# Quantum trajectories
# and Cushing's historical contingency

**Adriano Orefice***, **Raffaele Giovanelli** and **Domenico Ditto**
*Università degli Studi di Milano - DISAA*
*Via G. Celoria, 2 - 20133 - Milan (Italy)*

**Abstract -** With an apparent delay of over one century with respect to the development of standard Analytical Mechanics, but still in fully classical terms, the behavior of classical monochromatic wave beams in stationary media is shown to be ruled by a dispersive "Wave Potential" function, encoded in the structure of the Helmholtz equation. An exact, ray-based Hamiltonian description, revealing a strong ray coupling due to the Wave Potential, and reducing to the geometrical optics approximation when this function is neglected, is shown to hold even for typically wave-like phenomena such as diffraction and interference.

Recalling, then, that the time-independent Schrödinger equation (associating the quantum motion of mono-energetic particles with stationary monochromatic matter waves) is itself a Helmholtz-like equation, the mathematical treatment holding in the classical case is extended, without resorting to statistical concepts, to the exact, trajectory-based, Hamiltonian quantum dynamics of point-like particles. The particle trajectories and motion laws turn out to be coupled, in this case, by a function strictly analogous to the Wave Potential and formally assuming the familiar form of Bohm's "Quantum Potential", which is therefore not so much a "quantum" as a "wave" property - in whose absence the quantum particle dynamics reduces to the classical one. The time-independent Schrödinger equation is argued to be not a trivial particular case of the time-dependent one, but the exact quantum dynamical ground on which Schrödinger's time-dependent statistical description (representing particles as travelling wave-packets) is based. It provides indeed the (exact) link between classical particle dynamics and Bohm's hydrodynamics.

1- **Introduction**

The title of the present paper is suggested by Cushing's book **[1]**, reconstructing the historical contingency which favored the "Copenhagen hegemony", with particular regard to von Neumann's theorem **[2]**, excluding, even in principle, the possibility of "hidden variables" treatments. As Bell **[3]** tells it, "*I relegated the question to the back of my mind, and got on with more practical things*", until "*in 1951 I saw the impossible done. It was in papers by David Bohm*" **[4]**. As reported by Born **[5]**, however, Einstein himself - not to speak of the attitude of many other "founding fathers" of quantum physics - defined Bohm's approach "*too cheap*" for him, thus strongly contributing to push it into the realm of puzzling curiosities and to its long-lasting refusal, which did not restrain Bohm from maintaining and extending, together with a few co-workers **[6-8]**, his standpoint until his death. After this date his theory was kept alive by a somewhat exoteric line of thought **[9]**, and finally emerged to a quite generally recognized official life during the last decade **[10-21]**, mainly because of its applications to chemical physics and nano-technology.

---

* Corresponding author - adriano.orefice@unimi.it

The present (self-contained) work is the third of a series of papers **[22-23]** based on the observation that Bohm's approach to quantum particle motion is not "radical" enough, in the sense that the development of an exact, deterministic, trajectory-based *quantum dynamics* in terms of classical-looking, point-like particles is possible not only in principle, but in practice, without resorting to statistical wave-packets or dramatic conceptual changes with respect to classical dynamics.

The basic key for this development is the previous transition (performed here, in strictly *classical* terms, in Sect.2, with an apparent delay of over one century with respect to the standard Analytical Mechanics of wave-carrying media) from the *approximate and limited* eikonal ray description of *classical* monochromatic waves to their *exact*, but still trajectory-based, Hamiltonian treatment, where the ray trajectories turn out to be mutually coupled by a *strongly dispersive* "Wave Potential" function, encoded in the structure itself of the Helmholtz equation.

Starting, then, from de Broglie's and Schrödinger's well-known suggestion **[24-26]** that *classical particle dynamics* be the eikonal approximation of a more general *wave mechanics*, and from the consequent construction of the Helmholtz-like **time-independent** Schrödinger equation (associating the motion of mono-energetic particles with stationary monochromatic matter waves), the Wave Potential turns out to assume the form of Bohm's "Quantum Potential", which is therefore not so much a "quantum" as a "wave" property. Such a mono-energetic Quantum Potential is endowed, of course, with the same dispersive properties of the monochromatic Wave Potential, with which it shares the property of being the one and only cause of typically wave-like phenomena such as diffraction and/or interference.

Bohm's approach, on the other hand - starting from the **time-dependent** Schrödinger equation and representing the diffusive evolution of statistical wave-packets along a set of fluid-like flow lines according to a weighted average over the mono-energetic trajectories, doesn't differ so much from the standard Copenhagen paradigm. Although a time-dependent form (Bohm's original one, indeed) of the Quantum Potential may still be defined, the dispersive properties of the mono-energetic Quantum Potential are smoothed in this approach by its statistical nature. The **time-independent** Schrödinger equation is not, therefore, a trivial particular case of the **time-dependent** one, but the exact dynamical ground on which Bohm's statistical description is based. It's the exact dynamical link, indeed, between classical particle motion and Bohm's quantum hydrodynamics.

## 2 The case of classical waves

We assume here both *wave mono-chromaticity* and *stationary media*, allowing the best theoretical and experimental analysis of diffraction and/or interference patterns. Although our considerations may be easily developed for many kinds of *classical* waves, we shall refer in the present Section, in order to fix ideas, to a classical *electromagnetic* wave beam travelling through a stationary, isotropic and (generally)

A. Orefice et al. – **Quantum trajectories and Cushing's historical contingency**A. Orefice et al. – **Quantum trajectories and Cushing's historical contingency**

inhomogeneous dielectric medium according to a scalar wave equation of the simple form **[27]**

$$\nabla^2 \psi - \frac{n^2}{c^2} \frac{\partial^2 \psi}{\partial t^2} = 0 \;, \tag{1}$$

where $\psi(x,y,z,t)$ represents any component of the electric and/or magnetic field, $n(x,y,z)$ is the (time independent) refractive index of the medium and $\nabla^2 \equiv \frac{\partial^2}{\partial x^2} + \frac{\partial^2}{\partial y^2} + \frac{\partial^2}{\partial z^2}$ . By assuming

$$\psi(x,y,z,t) = u(x,y,z) e^{-i\omega t} \;, \tag{2}$$

with obvious definition of $u(x,y,z)$ and $\omega$, we get the Helmholtz equation

$$\nabla^2 u + (n k_0)^2 u = 0 \;, \tag{3}$$

where $k_0 \equiv \frac{2\pi}{\lambda_0} = \frac{\omega}{c}$. Notice that, limiting here our considerations to the case of monochromatic waves, we did not explicitly mention (for simplicity sake) the possible dependence of $\psi$, $u$ and $n$ on $\omega$. If we now perform the quite general and well-known replacement

$$u(x,y,z) = R(x,y,z) e^{i \varphi(x,y,z)} \;, \tag{4}$$

with real $R(x,y,z)$ and $\varphi(x,y,z)$, and separate the real from the imaginary part, eq.(3) splits into the coupled system **[27]**

$$\begin{cases} (\vec{\nabla} \varphi)^2 - (nk_0)^2 = \dfrac{\nabla^2 R}{R} \\ \vec{\nabla} \cdot (R^2 \vec{\nabla} \varphi) = 0 \end{cases} \tag{5}$$

where $\vec{\nabla} \equiv \partial/\partial \vec{r} \equiv (\partial/\partial x, \partial/\partial y, \partial/\partial z)$ and $\vec{r} \equiv (x,y,z)$. The function $R(x,y,z)$ represents the amplitude distribution of the beam, with no intrinsically statistical meaning. The *second* of eqs. (5) expresses the constancy of the flux of the vector $R^2 \vec{\nabla} \varphi$ along any tube formed by the field lines of the *wave vector*

$$\vec{k} = \vec{\nabla} \varphi \;. \tag{6}$$

As far as the *first* of eqs.(5) is concerned, we multiply it, for convenience, by the constant factor $\dfrac{c}{2 k_0}$, thus obtaining, by means of eq.(6), the relation

$$D(\vec{r},\vec{k}) \equiv \frac{c}{2k_0}[k^2 - (n\,k_0)^2 - \frac{\nabla^2 R}{R}] = 0 \ , \tag{7}$$

whose differentiation

$$\frac{\partial D}{\partial \vec{r}} \cdot d\vec{r} + \frac{\partial D}{\partial \vec{k}} \cdot d\vec{k} = 0 \ , \tag{8}$$

with $\partial/\partial\vec{k} \equiv (\partial/\partial k_x, \partial/\partial k_y, \partial/\partial k_z)$, immediately provides a Hamiltonian ray-tracing system of the form

$$\begin{cases} \dfrac{d\vec{r}}{dt} = \dfrac{\partial D}{\partial \vec{k}} = \dfrac{c\,\vec{k}}{k_0} \\ \dfrac{d\vec{k}}{dt} = -\dfrac{\partial D}{\partial \vec{r}} = \vec{\nabla}\,[\dfrac{c\,k_0}{2}n^2(x,y,z) - W(x,y,z)] \end{cases} \tag{9}$$

where

$$W(x,y,z) = -\frac{c}{2k_0}\frac{\nabla^2 R}{R} \tag{10}$$

and a ray velocity $\vec{v}_{ray} = \dfrac{c\,\vec{k}}{k_0}$ is implicitly defined. It is easily seen that, as long as $k \equiv |\vec{k}| = k_0$, we'll have $v_{ray} \equiv |\vec{v}_{ray}| = c$. The function $W(x,y,z)$, which we define in eq. (10) and call "Helmholtz Wave Potential", couples the rays of the beam in a kind of self refraction, strongly affecting their propagation. Such a term (which has the dimensions of a *frequency*) represents an intrinsic property encoded in the Helmholtz equation itself, and is determined by the structure of the beam. We observe, from the second of eqs.(5), that

$$\vec{\nabla} \cdot (R^2\,\vec{\nabla}\varphi) \equiv 2R\,\vec{\nabla} R \cdot \vec{\nabla}\varphi + R^2\,\vec{\nabla} \cdot \vec{\nabla}\varphi = 0. \tag{11}$$

This equations has a double role:
•*On the one hand*, since no new trajectory may suddenly arise in the space region spanned by the beam, we must have $\vec{\nabla} \cdot \vec{\nabla}\varphi = 0$, so that $\vec{\nabla} R \cdot \vec{\nabla}\varphi = 0$: the amplitude R is distributed at any time (together with its functions and derivatives) on the wavefront reached at that time, and both $\vec{\nabla} R$ and $\vec{\nabla} W$ are perpendicular to $\vec{k} \equiv \vec{\nabla}\varphi$. In other words, the term $\vec{\nabla} W$ acts *perpendicularly* to the Hamiltonian trajectories (9).
A basic consequence of this *general* property is the fact that, in the case of electromagnetic waves propagating *in vacuo*, the *absolute value* of the ray velocity remains equal to c all along each ray trajectory, because such a perpendicular term may only modify the *direction*, but not the *amplitude*, of the wave vector $\vec{k}$. The only possible changes of $k \equiv |\vec{k}|$ may be due, in a medium different from vacuum, to its refractive index *n(x,y,z)*, *but not to the action of the Wave Potential*.

• *On the other hand*, thanks to the constancy of the flux of $R^2 \vec{\nabla} \varphi$, the function *R(x,y,z)*, once assigned on the surface from which the beam is assumed to start, may be built up step by step, together with the Wave Potential *W(x,y,z)*, along the ray trajectories. The knowledge of the distribution of R on a wave-front is the necessary and sufficient condition to determine its distribution on the next wave-front. The Hamiltonian system (9) is "closed", in other words, by the second of eqs.(5). This allows its numerical integration, and provides **both** an exact *stationary* "weft" of coupled "rails" (which we could call "Helmholtz trajectories") along which the rays are channeled, **and** the ray motion laws along them, starting (with an assigned wave-vector) from a definite point of the launching surface and coupled by the Wave Potential *W(x,y,z)*.

Let us observe that when, in particular, the space variation length *L* of the beam amplitude *R(x,y,z)* satisfies the condition $k_0 L \gg 1$, the **first** of eqs.(5) is well approximated by the *eikonal equation* **[27]**

$$(\vec{\nabla} \varphi)^2 \simeq (nk_0)^2 \quad , \qquad (12)$$

decoupled from the **second** equation, and the term containing the wave potential *W(x,y,z)* may be dropped from the ray tracing system (9). In this *eikonal* (or "*geometrical optics*") *approximation* the rays are not coupled by the Wave Potential, and propagate independently from one another. The main consequence of this is the complete absence of such typically wave-like phenomena as diffraction and/or interference, which are therefore entirely due to the Wave Potential itself.

Coming back to the most general case, let us finally observe that if we pass to *dimensionless variables* by expressing:

• the space variable $\vec{r}$ (together with the space operators $\vec{\nabla}$ and $\nabla^2$) in terms of an "*a priori*" arbitrary physical length $w_0$ (such as the half-width of the slits in diffraction/interference cases),

• the wave vector $\vec{k}$ in terms of $k_0$, and

• the time variable *t* in terms of $w_0/c$,

and maintaining for simplicity the names $\vec{r}, \vec{k}, t$, the Hamiltonian system (9) takes on the dimensionless form

$$\begin{cases} \dfrac{d\vec{r}}{dt} = \vec{k} \\ \dfrac{d\vec{k}}{dt} = \dfrac{1}{2} \vec{\nabla} \left[ n^2 + \left(\dfrac{\varepsilon}{2\pi}\right)^2 G(x,y,z) \right] \end{cases} \qquad (13)$$

where

$$\varepsilon = \lambda_0 / w_0 \qquad (14)$$

and the Wave Potential (with opposite sign) is represented by the (dimensionless) function

$$G(x,y,z) = \frac{\nabla^2 R}{R} \quad . \tag{15}$$

Notice that different values of $\varepsilon \equiv \lambda_0/w_0$ (i.e. *different frequencies* $\omega = 2\pi c/\lambda_0$, for a fixed value of the assumed unit of length, $w_0$) lead to different values of the coefficient weighting the effect of the potential function G, and therefore to *different trajectories*. In this sense we may speak of a *dispersive* character of the *Wave Potential* itself. For a fixed value of $w_0$, waves of different frequencies travel along different trajectories, to which the coupling action of the Wave Potential maintains itself perpendicular.

Let us recall here that while our equations (9) provide an *exact* Hamiltonian description of wave trajectories, an *approximate* Hamiltonian description was presented in 1993/94 by one of the Authors (A.O., **[28, 29]**), for the propagation of electromagnetic Gaussian beams at the electron-cyclotron frequency in the magnetized plasmas of Tokamaks such as JET and FTU, and applied in recent years by an *équipe* working on the Doppler backscattering microwave diagnostics installed on the Tokamak TORE SUPRA of Cadarache **[30]**. A complex eikonal equation, amounting to a first order approximation of the beam diffraction, was adopted in **[28-30]** in order to overcome the collapse, for narrow wave beams, of the ordinary eikonal approximation.

**3- The case of quantum (matter) waves**

Let us pass now to the case of a mono-energetic beam of non-interacting particles of mass *m* launched with an initial momentum $\vec{p}_0$ into a force field deriving from a potential energy *V(x,y,z)* not explicitly depending on time. The *classical* motion of each particle of the beam may be described, as is well known, by the time-independent Hamilton-Jacobi equation **[27]**

$$(\vec{\nabla} S)^2 = 2m\,[E - V(x,y,z)] \quad , \tag{16}$$

where $E = p_0^2/2m$ is the total energy of the particle, and the basic property of the function *S(x,y,z)* is that the particle momentum is given by

$$\vec{p} = \vec{\nabla} S \quad . \tag{17}$$

The analogy between eqs.(16) and (17), on the one hand, and eqs.(12) and (6), on the other, together with an illuminating comparison between Fermat's and Maupertuis' variational principles, suggested to de Broglie **[24]** and Schrödinger **[25, 26]**, as is well known **[31]**, that the *classical* particle dynamics could be the *geometrical optics approximation* of a more general *wave-like reality* described by a suitable *Helmholtz-like* equation. Such an equation is immediately obtained, indeed, from eq.(3) by means of the replacements

$$\begin{cases} \varphi = \dfrac{S}{a} \\ \vec{k} \equiv \vec{\nabla}\varphi = \dfrac{\vec{\nabla}S}{a} = \dfrac{\vec{p}}{a} \\ k_0 \equiv \dfrac{2\pi}{\lambda_0} = \dfrac{p_0}{a} \equiv \dfrac{\sqrt{2mE}}{a} \\ n^2(x,y,z) = 1 - \dfrac{V(x,y,z)}{E} \end{cases} \quad (18)$$

directly inspired by the afore-mentioned analogy. The parameter "*a*" represents a constant *action* whose value is *a priori* arbitrary, but whose choice

$$a = \hbar \cong 1.0546 \times 10^{-27} \, erg \times s \quad (19)$$

is suggested by the de Broglie's Ansatz **[24]**

$$\vec{p} = \hbar \vec{k} \, , \quad (20)$$

thus transforming eq.(3) into the standard *time-independent* Schrödinger equation holding in a stationary field $V(x,y,z)$

$$\nabla^2 u + \dfrac{2m}{\hbar^2}[E - V(x,y,z)] \, u = 0 \, . \quad (21)$$

By applying now to the *Helmholtz-like* eq.(21) the same procedure leading from the Helmholtz eq.(3) to eqs.(5), making use of the first of eqs.(18) and assuming therefore

$$u(x,y,z) = R(x,y,z) \, e^{i S(x,y,z)/\hbar} \, , \quad (22)$$

eq.(21) splits into the coupled system **[32]**

$$\begin{cases} (\vec{\nabla}S)^2 - 2m(E - V) = \hbar^2 \dfrac{\nabla^2 R}{R} \\ \vec{\nabla} \cdot (R^2 \vec{\nabla}S) = 0 \end{cases} \, , \quad (23)$$

analogous to eqs.(5). By simply maintaining eq.(17), the first of eqs. (23) may be written in the form of a generalized, time-independent Hamiltonian

$$H(\vec{r}, \vec{p}) \equiv \dfrac{p^2}{2m} + V(x,y,z) + Q(x,y,z) = E \quad (24)$$

where the function

$$Q(x,y,z) = -\dfrac{\hbar^2}{2m} \dfrac{\nabla^2 R}{R} \quad (25)$$

(which has the dimensions of an *energy*) is structurally analogous to the Wave Potential function *W(x,y,z)* of eq.(10), and turns out to formally coïncide with the well known *Quantum Potential* of Bohm's theory**.** Such a term is clearly due not so much to the "quantum" behavior of the particles as to their "wave-like" nature, suggested by de Broglie and Schrödinger. By differentiating eq. (24) we get the relation

$$\frac{\partial H}{\partial \vec{r}} \cdot d\vec{r} + \frac{\partial H}{\partial \vec{p}} \cdot d\vec{p} = 0 \qquad (26)$$

with $\partial/\partial \vec{p} \equiv (\partial/\partial p_x, \partial/\partial p_y, \partial/\partial p_z)$, leading to a Hamiltonian dynamical system of the form

$$\begin{cases} \dfrac{d\vec{r}}{dt} = \dfrac{\partial H}{\partial \vec{p}} = \dfrac{\vec{p}}{m} \\ \dfrac{d\vec{p}}{dt} = -\dfrac{\partial H}{\partial \vec{r}} = -\vec{\nabla}[V(x,y,z) + Q(x,y,z)] \end{cases} \qquad (27)$$

The Hamiltonian treatment is allowed by our mono-energetic, time independent approach, such that *Q(x,y,z)*, just like *V(x,y,z)*, is a stationary, merely geometric function **[33]**.

This *quantum* dynamical system is strictly similar to the exact, deterministic ray-tracing system (9) concerning *classical* electromagnetism. In spite of its "quantum" context, therefore, we shall submit it to the same interpretation and mathematical treatment applied in the previous (classical) case. Once more, the function *R(x,y,z)* shall be assumed to represent the amplitude distribution of a beam, with no intrinsically statistical meaning. The presence of the potential *Q(x,y,z)* causes, in its turn, the "*Helmholtz coupling*" of the beam trajectories, and its absence or omission would reduce the *quantum* system (27) to the standard *classical* set of dynamical equations, which constitute therefore, as expected **[24-26]**, its *geometrical optics approximation*.

In complete analogy with the *classical* electromagnetic case of the previous Section,

1) the term $-\vec{\nabla} Q(x,y,z)$ (behaving here as a force) is perpendicular to $\vec{p} \equiv \vec{\nabla} S$, so that it cannot modify the *amplitude* of the particle momentum (while modifying, in general, its *direction*), and the only possible amplitude changes of $\vec{p}$ could be due to the presence of an external potential *V(x,y,z)*: in other words, *no energy exchange may ever occur between particles and Quantum Potential*;

2) the relations $\vec{p} = \vec{\nabla} S$ and $\vec{\nabla} \cdot (R^2 \vec{\nabla} S) = 0$ allow to obtain step by step, along the particle trajectories, both R(x,y,z) and Q(x,y,z), thus "closing" the Hamiltonian system (27) and providing the exact, complete, deterministic dynamics of *classical-looking, point-like particles* starting from assigned point-like positions, and following well defined *stationary* trajectories coupled by the Quantum Potential.

Each (point-like) particle of the beam is endowed, at any time, with a well defined momentum, associated with its instantaneous (point-like) position. *No wave-packet and no statistical representation are employed in this monochromatic "hidden variables" treatment.*

In complete analogy, moreover, with the previous electromagnetic case, the quantum Hamiltonian system (27) may be put in a suggestive *dimensionless* form by expressing lengths (as well as $\vec{\nabla}$ and $\nabla^2$) in terms of a physical length $w_0$ ( to be defined later on), momentum in terms of $p_0$ and time in terms of $w_0/v_0$, with $v_0 = p_0/m$:

$$\begin{cases} \dfrac{d\vec{r}}{dt} = \vec{p} \\ \dfrac{d\vec{p}}{dt} = \dfrac{1}{2}\vec{\nabla}\,[\,-\dfrac{V}{E} + (\dfrac{\varepsilon}{2\pi})^2\ G(x,y,z)] \end{cases} \quad (28)$$

where the parameter $\varepsilon$ and the (dimensionless) potential function $G(x,y,z)$ are given, once more, by eqs.(14) and (15). Not surprisingly, the *quantum* system (28) turns out to formally coïncide with the *classical* dimensionless system (13) by simply replacing $\vec{k}$ by $\vec{p}$ and $n^2$ by *(1-V/E)*, in agreement with eqs.(18).

The trajectory coupling due to *G(x,y,z)* is therefore a physical phenomenon affecting *both classical and quantum waves*, and its absence would reduce the relevant equations to the ones, respectively, of standard geometrical optics and of classical dynamics.

Let us observe once more that different values of $\varepsilon \equiv \lambda_0/w_0$ (i.e. different total energy *E*, for a fixed value of the assumed unit of length, $w_0$) lead to different sets of trajectories, i.e. to a dispersive behavior.

**4- Numerical examples**

Once assigned on the launching surface of the beam, the wave amplitude profile *R(x,y,z)* and the consequent potential function *G( x,y,z )* may be numerically built up step by step, together with their derivatives, along the beam trajectories. We present here some applications of the Hamiltonian systems (13) and/or (28) to the propagation of collimated beams injected at *z = 0*, parallel to the *z*-axis, simulating wave *diffraction* and/or *interference* through suitable slits, each one of half width $w_0$. *Here we perform, therefore, the choice of the physical meaning of this length*, and we assume $\varepsilon \equiv \lambda_0/w_0 < 1$.

The problem is faced by taking into account, for simplicity sake, either (*quantum*) particle beams in the absence of external fields (*V = 0*) or (*classical*) electromagnetic beams *in vacuo* $(n^2 = 1)$, with a geometry allowing to limit the computation to the *(x,z)*-plane. Because of the coincidence between the (dimensionless) Hamiltonian systems (13) and (28), the only choice to be performed is between the variable names $\vec{k}$ or $\vec{p}$ - and we opt here for the second one, reminding however that we are not necessarily speaking of quantum topics. Recalling that, because of the transverse nature of the gradient $\vec{\nabla}G$, the *amplitude* of $\vec{p}$ remains unchanged (in the absence of external fields and/or refractive effects) along each trajectory, we have

$$\begin{cases} p_x(t=0) = 0; \ p_z(t=0) = 1 \\ p_z(t \geq 0) = \sqrt{1 - p_x^2(t \geq 0)} \end{cases} \quad (29)$$

and the dimensionless Hamiltonian system (28) reduces to the form

$$\begin{cases} \dfrac{dx}{dt} = p_x \\ \dfrac{dz}{dt} = \sqrt{1 - p_x^2} \\ \dfrac{dp_x}{dt} = \dfrac{\varepsilon^2}{8\pi^2} \dfrac{\partial G(x,z)}{\partial x} \end{cases} \quad (30)$$

where

$$G(x,z) \equiv \frac{\nabla^2 R}{R} = \frac{\partial^2 R / \partial x^2}{p_z^2 \, R} \quad (31)$$

and the Hamiltonian system (30) is "closed" by the second of eqs. (23). Considering, in the present case, a set of rays labelled by the index $j$, if:

$x_j(t)$ and $z_j(t)$ are the coordinates of the point reached by the $j$-th ray at the time $t$;

$R_j(t)$ is the value assumed by the wave amplitude along such a ray at the time $t$;

and

$$d_j(t) = \sqrt{(x_j(t) - x_{j-1}(t))^2 + (z_j(t) - z_{j-1}(t))^2} \quad (32)$$

is the distance between two adjacent rays at the same time step, the "closure" equation may be written in the form

$$R_j^2(t) d_j(t) = const. \quad (33)$$

We assumed throughout the present computations the value $\varepsilon \equiv \lambda_0 / w_0 = 1.65 \times 10^{-4}$. Let us mention, for comparison, that a case of cold neutron diffraction was considered in **Ref.[11]** with

$$\begin{aligned} \lambda_0 &= 19.26 \times 10^{-4} \, \mu m, \ 2w_0 = 23 \, \mu m, \\ \varepsilon &= \lambda_0 / w_0 \cong 1.67 \times 10^{-4} \end{aligned} \quad (34)$$

The beam launching amplitude distribution $R(x; z = 0)$ (from whose normalization the function $G$ is obviously independent) is assigned, in the following, by means of two different models consisting of suitable superpositions of Gaussian functions either in the form

$$R(x; z=0) = a \exp(-q^2 x^2) + b \sum_{N=1}^{M} \left\{ \exp\left[-q^2 (x - N x_C)^2\right] + \exp\left[-q^2 (x + N x_C)^2\right] \right\} \quad (35a)$$

or in the form

A. Orefice et al. – **Quantum trajectories and Cushing's historical contingency**

$$R(x;z=0) = \sum_{N=-M}^{M} \left\{ exp\left[-q^2\left(x - x_C + N x_1\right)^2\right] + exp\left[-q^2\left(x + x_C + N x_1\right)^2\right]\right\} \qquad (35b)$$

allowing a wide variety of beam profiles, and an arbitrary number of "slits", according to the choice of the parameters $a, b, q, M, x_C, x_1$.

The values of $R(x;z>0)$ are then computed step by step by means of a symplectic integration method, and connected, at each step, by a Lagrange interpolation, allowing to perform space derivatives and providing both $G(x;z>0)$ and the full set of trajectories. The fringeless diffraction case of a simple Gaussian beam is obtained from eq.(35a) with $a = q = 1; \ b = 0$, and is presented in Figs.1-3 showing, respectively, the initial and final transverse profiles of the beam intensity ($\div R^2(x,z)$, in arbitrary units) and of the potential function $G(x,z)$, and the corresponding set of trajectories on the $(z,x)$-plane. A complete numerical coincidence is evidenced in Fig.3 (heavy lines) with the analytical *waist trajectories*

$$x = \pm \sqrt{1 + (\frac{\varepsilon z}{\pi})^2} \qquad (36)$$

representing, *in vacuo*, the rays starting (at $z = 0$) from the "waist" positions $x = \pm 1$, which delimit the so-called "paraxial" part **[23,34]** of the Gaussian beam. This coincidence provides an excellent test of our approach and interpretation.

Let's observe, moreover, that the presence of the parameter $\varepsilon \equiv \lambda_0 / w_0$ in eq.(36) is a proof of the dispersive character of diffraction, and therefore of the Wave (or Quantum) Potential to which it is due.

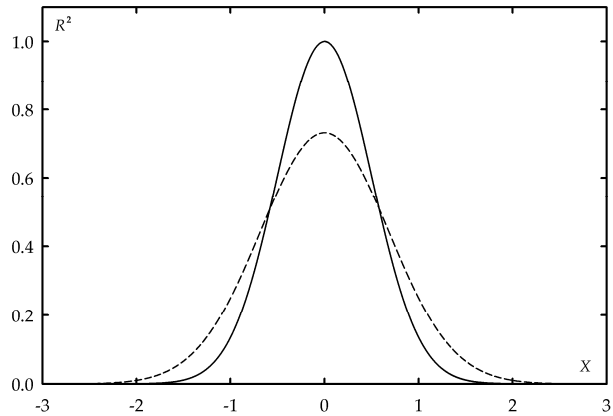

**Fig.1** Initial (continuous) and final (dashed) transverse intensity profiles for a Gaussian beam with $\varepsilon \equiv \lambda_0 / w_0 = 1.65 \times 10^{-4}$.

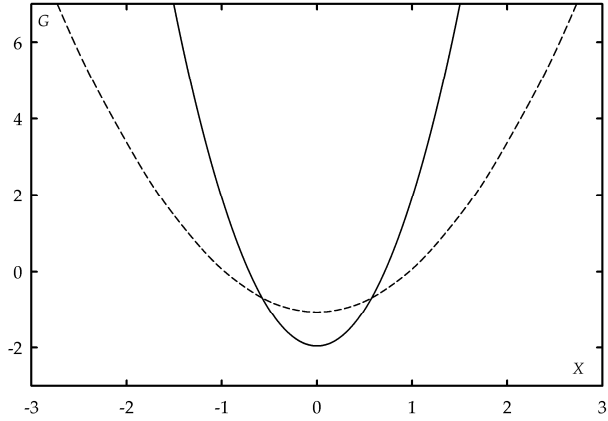

**Fig.2** Initial (continuous) and final (dashed) transverse profiles of the potential function G corresponding to Fig.1.

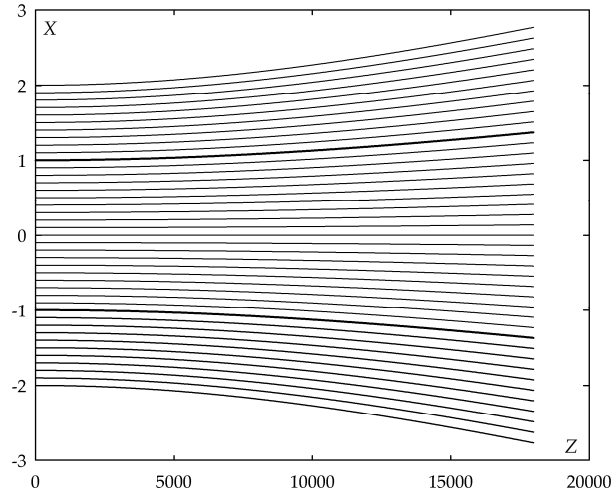

**Fig.3** Trajectories on the (z, x)-plane corresponding to Figs.1-2. The two heavy lines represent the paraxial waist lines of the beam.

It is also worthwhile recalling here that a simple consequence of eq.(36) is the fact **[23]** that a particle of a Gaussian beam entering a slit of half-width $w_0$, placed at $z=0$, with momentum components $p_x(t=0)=0$; $p_z(t=0)=1$, at an unknown position $x$ (i.e. with a space uncertainty $\Delta x = 2w_0$), acquires, because of the diffractive process (i.e. under the action of the Quantum Potential) an unknown transverse momentum ranging between $p_x \simeq \pm 4\hbar/(2w_0)$, so that

$$\Delta x\, \Delta p_x \simeq 8\hbar \quad , \qquad (37)$$

a suggestive relation providing a circumstantial evidence that uncertainty resides in our knowledge, and not in the intrinsic nature of physical reality.

We present in Fig.4 and Fig.5, respectively, the initial and final transverse profiles (showing a clear fringe formation) of the beam intensity $R^2$ and of the potential function $G$ for the single-slit diffraction case obtained from eq.(35a) with $a = 0$, $b = 1$, $q = 1.68$, $M = 2$, $x_C = 0.31$. Fig.6 shows, in its turn, the corresponding set of trajectories on the $(z,x)$-plane.

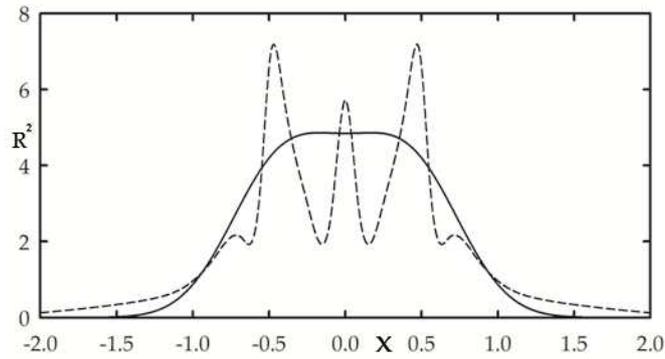

**Fig.4** Initial (continuous) and final (dashed) transverse profiles of the beam intensity for the diffraction-like case obtained from eq. (35a) with $a = 0, b = 1, q = 1.68, M = 2, x_C = 0.31$.

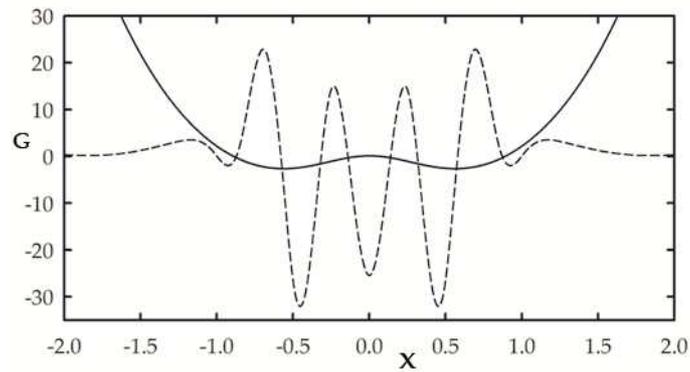

**Fig.5** Initial (continuous) and final (dashed) transverse profiles of the potential function G corresponding to Fig.4.

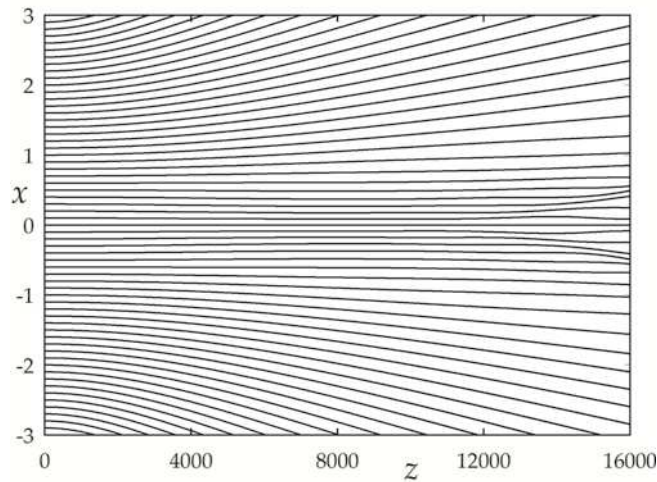

**Fig.6** Beam trajectories corresponding to Figs.4-5.

Finally Fig.7 and Fig.8 present, respectively, the initial (continuous) and final (dashed) transverse profiles of **beam intensity** and **potential function** $G(x,z)$ for the case obtained from eq.(35b) with $q = 3.5$; $M = 3$; $x_C = 1.15$; $x_1 = 0.3$, and Fig.9 shows the corresponding set of beam **trajectories** on the (z,x)-plane.

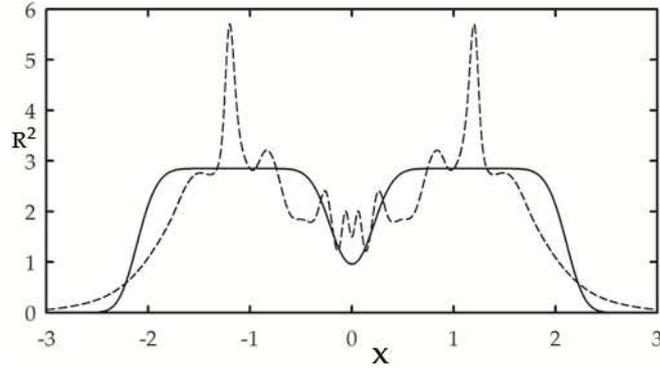

**Fig.7**  Initial (continuous) and final (dashed) transverse profiles of the **beam intensity** for the diffraction-like case obtained from eq.(42) with $q = 3.5$; $M = 3$; $x_C = 1.15$; $x_1 = 0.3$.

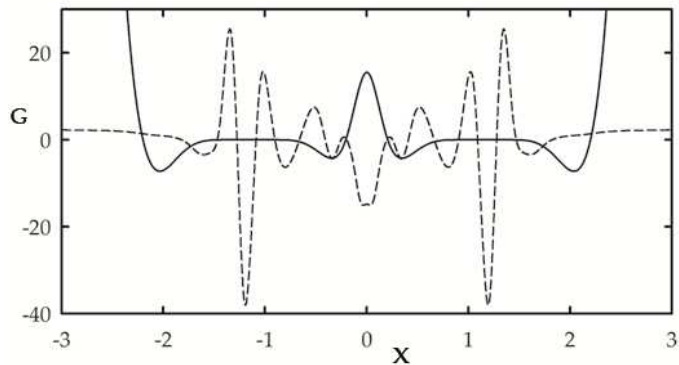

**Fig.8**  Initial (continuous) and final (dashed) transverse profiles of the potential function G corresponding to Fig.5.

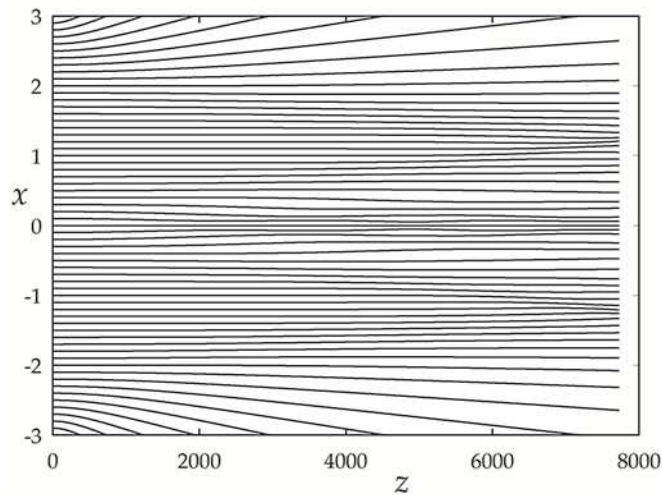

**Fig.9**  Beam trajectories corresponding to Fig.7.

**5- Discussion of Bohm's approach**

Starting from eqs.(2) and (21) it's a standard procedure to get

$$\nabla^2 \psi - \frac{2m}{\hbar^2} V(x,y,z)\,\psi = -\frac{2m}{\hbar^2} E\,\psi \equiv -\frac{2mi}{\hbar}\frac{E}{\hbar\omega}\frac{\partial \psi}{\partial t} \quad , \tag{38}$$

a relation which, by assuming the Planck relation

$$E = \hbar\omega, \tag{39}$$

i.e. by attributing to the energy of a material particle a relation coming, *stricto sensu*, from the radiation theory **[31]** - takes on the usual form of the time-dependent Schrödinger equation in a stationary potential field *V(x,y,z)*,

$$\nabla^2 \psi - \frac{2m}{\hbar^2} V(x,y,z)\,\psi = -\frac{2m\,i}{\hbar}\frac{\partial \psi}{\partial t} \quad , \tag{40}$$

where *E* and *ω* are not involved, and no wave dispersion is therefore, in principle, described. As is well known, indeed, eq.(40) itself (representing a rare example of an intrinsically complex diffusion-like equation in physics) is not even a wave equation.
We recall however that, just like the Helmholtz equation (3) is associated with the wave equation (1), the *Helmholtz-like* eq.(21) is associated - *via* eqs.(2) and (39) - with the ordinary-looking wave equation

$$\nabla^2 \psi = \frac{2m}{(\hbar\omega)^2}(E-V)\frac{\partial^2 \psi}{\partial t^2} \equiv \frac{2m}{E^2}(E-V)\frac{\partial^2 \psi}{\partial t^2} \tag{41}$$

providing significant information **[31]** about the wave propagation and the dispersive character of a mono-energetic particle beam. To be sure, the most important and reassuring information is the fact itself that mono-energetic quantum matter waves propagate exactly like all other physical waves.

While eq.(40) may be considered as a mathematical truism, its "stronger", but generally accepted, version:

$$\nabla^2 \psi - \frac{2m}{\hbar^2} V(x,y,z,t)\,\psi = -\frac{2m\,i}{\hbar}\frac{\partial \psi}{\partial t} \quad , \tag{42}$$

containing a *time dependent* potential *V(x,y,z,t)*, may only be assumed as a separate Ansatz, justified by various plausibility arguments (see, for instance, **Ref.[35]**) and by its current application in such fields as molecular **[36]** and ultra-fast laser **[37, 38]** dynamics.
Bohm's approach applies, as is well known **[4]**, a replacement of the form

A. Orefice et al. – **Quantum trajectories and Cushing's historical contingency**$$\psi(x,y,z,t) = R(x,y,z,t)\, e^{i\,S(x,y,z,t)/\hbar} \tag{43}$$

to eq.(42) itself, splitting it, after separation of real and imaginary parts, into the system

$$\begin{cases} \dfrac{\partial P}{\partial t} + \vec{\nabla}\cdot\left(P\dfrac{\vec{\nabla} S}{m}\right) = 0 \\ \dfrac{\partial S}{\partial t} + \dfrac{(\vec{\nabla} S)^2}{2m} + V(x,y,z,t) - \dfrac{\hbar^2}{2m}\dfrac{\vec{\nabla} R}{R} = 0 \end{cases} \tag{44}$$

In agreement with the standard Copenhagen interpretation, the function $P = R^2$ is assumed to represent, in Bohm's approach, the probability density for particles of different energies, belonging to a statistical ensemble. While however the first of eqs.(44) is viewed as a *fluid-like* continuity equation for such a density, the second one, having the form of a *dynamical* Hamilton-Jacobi equation including a "Quantum Potential" of the form

$$Q(x,y,z,t) = -\dfrac{\hbar^2}{2m}\dfrac{\nabla^2 R(x,y,z,t)}{R(x,y,z,t)}, \tag{45}$$

(to be compared with eq.(25)) is viewed as suggesting that "*precisely definable and continuously varying values of position and momentum*" **[4]** may be associated, in principle, to point-like particles. In spite of this, "*the most convenient way of obtaining R and S*", according to Bohm, "*is to solve [eq.(42)] for the Schrödinger wave function*", leading *de facto* to a statistical description of the particle motion in terms of wave-packets and hydrodynamic flow-lines. Let us notice that, in spite of the formal coincidence of eq.(45) with the stationary expression (25), monochromatic features due to dispersion and transverse trajectory coupling may only survive, in this case, as distorted averages.

The situation is much the same, and even more evident, if we limit our attention to a *stationary* external potential $V(x,y,z)$. In this case, as is well known **[31,32]**, the *time-independent* Schrödinger equation (21) admits in general a (discrete or continuous, according to the boundary conditions) set of orthonormal eigen-modes and of energy eigen-values, which, referring for simplicity to the discrete case, we shall call, respectively, $u_n(x,y,z)$ and $E_n$. If we make use of eqs.(2) and (39), and define the eigen-frequencies $\omega_n \equiv E_n/\hbar$, together with the eigen-waves

$$\psi_n(x,y,z,t) = u_n(x,y,z)\, e^{-i\omega_n t} \equiv u_n(x,y,z)\, e^{-i\frac{E_n}{\hbar} t} \tag{46}$$

and with an arbitrary linear superposition of them,

$$\psi(x,y,z,t) = \sum_n c_n\, \psi_n \tag{47}$$

(with constant coefficients $c_n$), such a superposition, when inserted into the *time dependent* Schrödinger equation (40), reduces it to the form

$$\sum_n c_n \, e^{-i\frac{E_n}{\hbar}t} \left\{ \nabla^2 u_n + \frac{2m}{\hbar^2}[E_n - V(x,y,z)] \, u_n \right\} = 0 \qquad (48)$$

showing that it provides (in duly normalized form) a general solution of eq.(40) itself. A solution whose Born interpretation **[39]**, even though "*no generally accepted derivation has been given to date*" **[40],** has become one of the main principles of quantum mechanics.

Let us notice once more that, in spite of the formal coincidence between the function Q and the expression (25), any monochromatic feature due dispersion and transverse trajectory coupling is mingled and smoothed here in an average behavior. Since, indeed, the function $R$ is given by a sum over the full set of eigen-functions, there is no possibility of distinguishing the peculiarities of the single monochromatic terms.

Once more Bohm's approach doesn't appear to differ so much from the standard Copenhagen paradigm, to which it merely associates a set of *fluid-like* trajectories, providing a weighted average over the *dynamical* mono-chromatic ones, which are simply *assumed* to exist. As suggested by Bohm himself, who presented the use of "*statistical ensembles as a practical necessity, but not as a manifestation of an inherent lack of determination of the particle nature and motion*" **[4],** the **time-dependent** Schrödinger equation describes *de facto* the progressive diffusive evolution of a statistical information assigned from the beginning, in the form of a wave packet, as a weighted average.

**6- Conclusions**

Recalling that in the stationary state form of the Hamilton-Jacobi formulation we have

$$\frac{\partial S}{\partial t} = -E \, , \qquad (49)$$

Bohm's equations (44) are seen to reduce, for particles of total energy *E,* to the simple system

$$\begin{cases} \vec{\nabla} \cdot (P \dfrac{\vec{\nabla} S}{m}) = 0 \\ -E + \dfrac{(\vec{\nabla} S)^2}{2m} + V(x,y,z) - \dfrac{\hbar^2}{2m} \dfrac{\vec{\nabla} R}{R} = 0 \end{cases} \qquad (50)$$

formally coinciding with our equation system (23) - although Bohm's statistical interpretation of the function $P = R^2$ makes an exact Hamiltonian treatment problematic. This statistical interpretation, to be sure, makes the basic difference.

While in fact the stationary mono-energetic case (50) could appear to be a particular and minor case of the statistical approach (44), this is not the case for eqs.(23). As we have shown, indeed, our Hamiltonian treatment provides, in terms of a dispersive Quantum Potential, the exact dynamical ground on which the quantum statistical description is based, and therefore the "missing link" between classical point-like particle dynamics and Bohm's statistical wave-packets.

The absence of this "link" may justify both Cushing's historical contingency and Einstein's disdainful attitude.